# Probing the Electronic States in Black Phosphorus Vertical Heterostructures

Xiaolong Chen[1], Lin Wang[2], Yingying Wu[1], Heng Gao[3], Yabei Wu[3], Guanhua Qin[3], Zefei Wu[1], Yu Han[1], Shuigang Xu[1], Tianyi Han[1], Weiguang Ye[1], Jiangxiazi Lin[1], Gen Long[1], Yuheng He[1], Yuan Cai[1], Wei Ren[3], Ning Wang*,[1]

[1] *Department of Physics and the William Mong Institute of Nano Science and Technology, the Hong Kong University of Science and Technology, Hong Kong, China*

[2] *Department of Condensed Matter Physics, Group of Applied Physics, University of Geneva, 24 Quai Ernest Ansermet, CH1211 Geneva, Switzerland*

[3] *International Centre for Quantum and Molecular Structures, Materials Genome Institute, and Department of Physics, Shanghai University, Shanghai, China*

**E-mail**: phwang@ust.hk

**Abstract**
Atomically thin black phosphorus (BP) is a promising two-dimensional material for fabricating electronic and optoelectronic nano-devices with high mobility and tunable bandgap structures. However, the charge-carrier mobility in few-layer phosphorene (monolayer BP) is mainly limited by the presence of impurities or disorder structures. In this study, we demonstrate that vertical BP heterostructure devices offer great advantages in probing the electron states of monolayer and few-layer phosphorene at temperatures down to 2 K through capacitance spectroscopy. Electronic states in the conduction and valence bands of phosphorene are accessible over a wide range of temperature and frequency. Exponential band tails have been determined to be related to disorders. Unusual phenomena such as the giant temperature-dependence of the electron state population in few-layer phosphorene have been observed and systematically studied. By combining the first-principles calculation, we identified that the thermal excitation of charge trap states and oxidation-induced defect states were the main reasons for the giant temperature dependence of the electron state population and degradation of the on-off ratio in phosphorene field-effect transistors.

## 1. Introduction

Atomically thin black phosphorus (BP), a two-dimensional layered material held together by van der Waals interactions, has attracted great attentions recently owing to its unique properties[1-11]. Few-layer phosphorene normally has high charge-carrier mobility[1, 9, 10] (~ $1000 \text{ cm}^2\text{V}^{-1}\text{s}^{-1}$) in comparison with transition metal dichalcogenides at room temperature[12-15]. The strong ambipolar field effects[1-3, 16, 17] and tunable bandgap[16-20] observed in phosphorene allow the realization of electrically tunable PN junctions[5], heterojunctions[21] and high-performance radio-frequency transistors[22]. Thanks to its small and direct bandgap[18, 19], few-layer phosphorene also shows fast photo-response and broad-band energy harvesting up to the near-infrared[4, 5, 23] range.

To further boost up the applications of BP, its charge-carrier mobility needs to be optimized. Unfortunately, the carrier mobility of atomically thin BP is still much lower than the theoretically predicted value (~ $10000 \text{ cm}^2\text{V}^{-1}\text{s}^{-1}$ [24]) due to the presence of disorders and impurities. The preparation of monolayer and few-layer phosphorene (thickness < 2 nm) is still challenging because phosphorene is easily oxidized in the atmosphere[1-3, 25, 26]. Investigation of the electron states of atomically thin BP is important for better understanding the relationship between charge carrier behavior, band structure and impurity states in BP and other types of two-dimensional (2D) semiconductors[15, 27-29]. In this study, we demonstrate the advantages of capacitance spectroscopy techniques for investigating the electronic states in a vertical BP heterostructure (Fig. 1a-c) which is different from conventional field-effect-transistor (FET) devices. In conventional 2D semiconductor FET structures, charge carriers normally suffer from a large lateral resistance and electrons are often localized near the band edge, causing difficulties in detecting electronic states at low temperature and high frequency[15, 27, 28] conditions. In contrast, the electrons in the vertical configuration reported in our previously work[15] can respond directly and efficiently (Fig. 1b) without suffering the large lateral resistance near the band edge. We show that the electron states in both monolayer and few-layer phosphorene can be easily detected over wide ranges of temperature (2 K-300 K) and frequency (100 Hz-1 MHz). A number of interesting properties in few-layer phosphorene such as the giant variation of electron state population at different temperatures, exponential band tail changes for different Fermi energies, thickness-dependent band gap width and variation of charge trap density have been observed.

## 2. Results and discussion

As shown in Fig. 1a-c, the vertical heterostructure consists of a thin BP flake and a hexagonal boron nitride (hBN) dielectric layer (6 nm-20 nm) sandwiched by a bottom gate (Cr/Au, 2 nm/25 nm) supported on a 300 nm $SiO_2$/p-Si substrate and a top electrode (Ti/Au, 5 nm/50 nm). Ultrathin BP flakes are mechanically exfoliated and identified in a glove box equipped with an optical microscope in nitrogen atmosphere in order to avoid sample quality degradation. The fully covered top electrode further isolates BP flakes from oxidation when transferring the sample through air. By applying an alternating current (AC) parameterized with a frequency $f$ and a gate voltage $V_g$ to the vertical structure (Fig. 1c), electron states of BP are excited and contribute to the capacitance $C_{BP}$ in a series connection with the geometric capacitance $C_g$ (Fig.

1d). Then the measured capacitance is expressed as $C_t = (C_g^{-1} + C_{BP}^{-1})$ after excluding the parallel residue capacitance $C_p$ (see details in the Supporting Information).

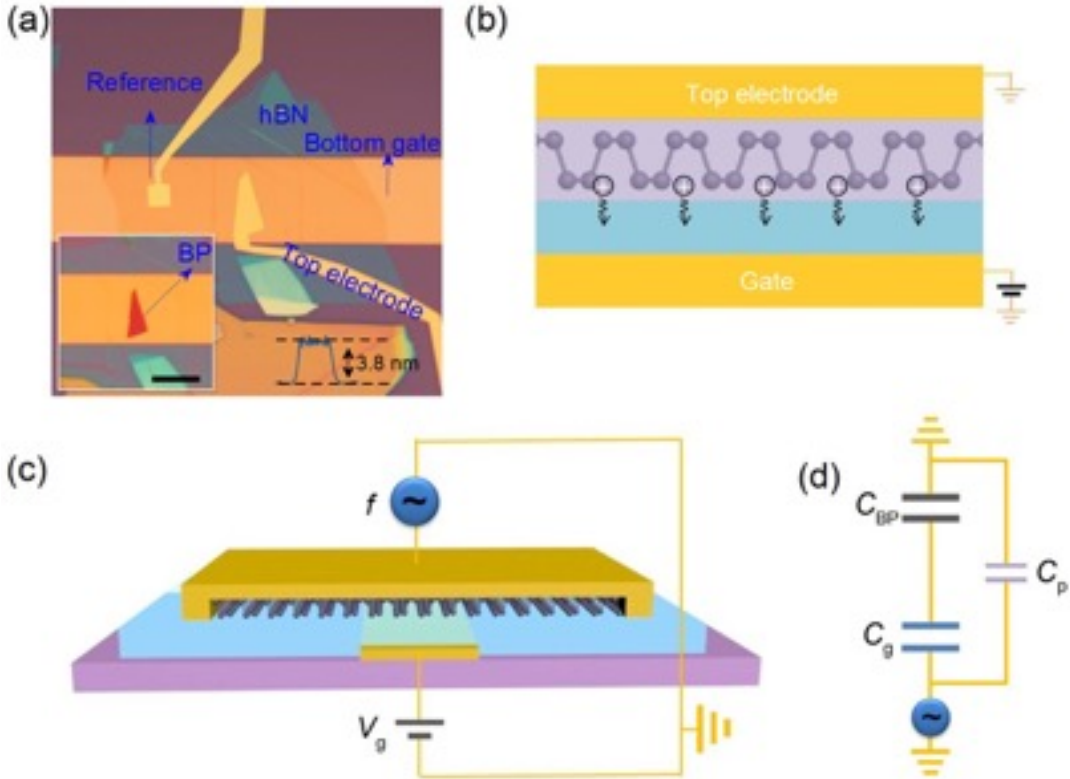

**Figure 1.** The configuration of BP vertical heterostructure. (a) Optical image of the BP vertical heterostructure. The inset shows the 3.8nm-thick BP on hBN/bottom gate before covering the top electrode. The sample thickness is determined by an atomic force microscope. The scale bar is 10 μm. (b, c) Schematic images of the BP vertical heterostructure, in which electrons respond vertically without suffering the large lateral resistance. (d) The equivalent circuit of the BP capacitance devices.

Fig. 2d illustrates the capacitance data measured as a function of gate voltage for a 7.2 nm-thick BP sample at 2 K. The strong ambipolar capacitance behavior indicates that both hole and electron doping states have been achieved. The top electrode has an excellent Ohmic contact with the valence band of the BP sample and the Schottky barrier to the conduction band is overcome when the excitation voltage ($\geq 100$ meV) is applied at 100 kHz (see the inset of Fig. 2d). When the gate voltage is reversed to be negative, holes start to accumulate at the BP surface (Fig. 2a), while electrons form at the interface when gate voltage $V_g > 0$ (Fig. 2c). In both cases, if $|V_g|$ is sufficiently large, the measured capacitance would approach the geometry capacitance $C_g$ (denoted by the blue dashed line in Fig. 2d). When the carriers are depleted (Fig. 2b), the measured capacitance reaches the minimum value $C_{min} = (C_g^{-1} + \frac{d_{BP}}{\varepsilon_{BP}})^{-1}$, where $\varepsilon_{BP}$ and $d_{BP}$ are the dielectric constant and thickness of BP sample respectively. Inside the bandgap, however, the measured capacitance $C_{BP}$ in this vertical configuration is normally non-zero because of the bulk capacitance $C_{bulk} = \frac{\varepsilon_{BP}}{d_{BP}}$ in parallel connection with the quantum capacitance of BP $C_q$, which is known to be proportional to the density of states (DOS)[15, 27, 30-33]. Then the quantum capacitance $C_q$ can be approximately expressed by $C_q = C_{BP} - \frac{\varepsilon_{BP}}{d_{BP}}$. Without considering thermal effects, the DOS of the sample can be conveniently expressed by $DOS = \frac{C_q}{e^2}$ [31-33]. The theoretical simulation of these unique characteristics of the vertical structure is demonstrated in the Supporting Information.

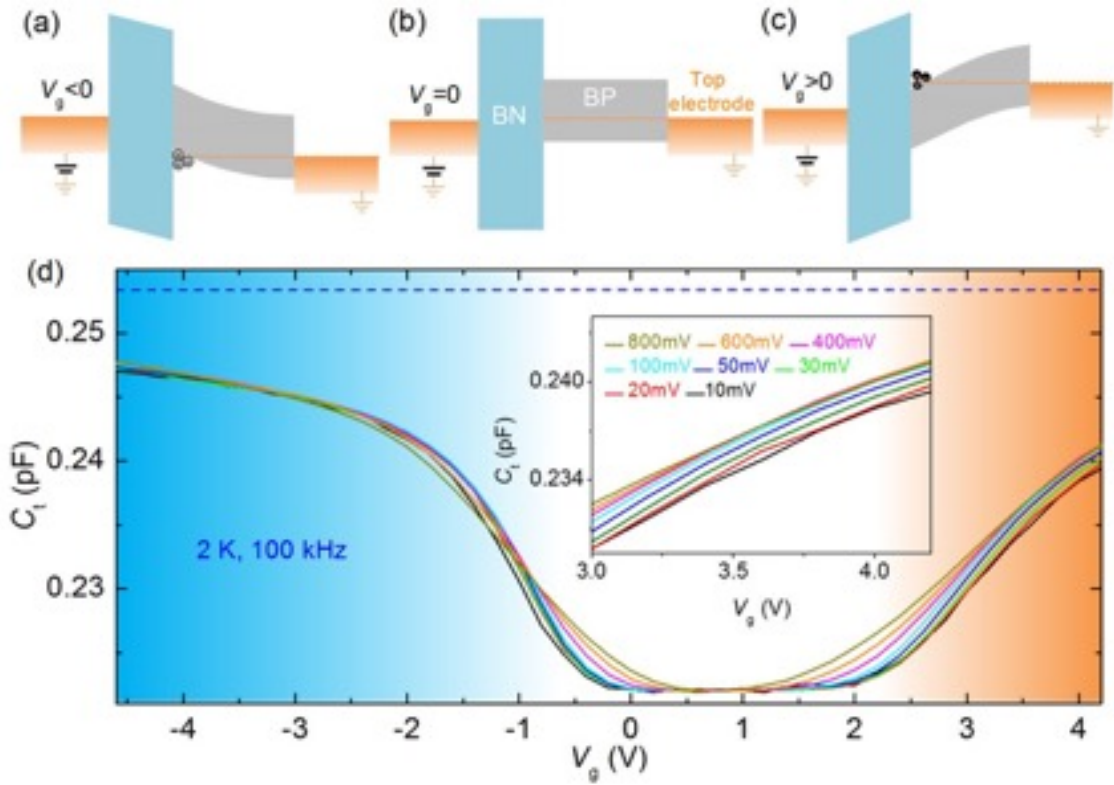

**Figure 2.** The band structure and capacitance characteristics of the BP vertical heterostructure. (a-c) Schematic band diagrams in the regions for hole accumulation (a), depletion (b) and electron accumulation (c). (d) Measured total capacitance $C_t$ as a function of gate voltage $V_g$ for the 7.2 nm-thick BP sample at $T = 2$ K and $f = 100$ kHz with different excitation voltages. The dashed line indicates the geometry capacitance $C_g$. The inset shows the details of the $C_t$ near the conductance edge.

For both monolayer and few-layer phosphorene samples, the electron states in valence and conduction bands are accessible through the vertical heterostructure configurations. As shown in Fig. 4a-d, the measured capacitance generally shows a clear ambipolar behavior from room temperature down to 2 K. This allows us to directly determine the band gap width of BP at low temperatures (≤ 30 K) where the thermal excitation of free carriers and charge trap states are suppressed. In order to precisely determine the band gap width, the band tail (due to the presence of impurities and disorders) should be considered. The exponential characteristic of the DOS[27] in our sample (excluding charge trap states $D_{it}$) matches the band tail's form, $DOS \propto \exp(\frac{|E|-|E_b|}{E_U})$, where $E_{bv} \leq E \leq E_{bc}$ and $E_U$ characterizes the energy width of the band tail. $E_{bc}$ and $E_{bv}$ are the positions of conduction and valence bands respectively.

As shown in Fig. 3b, the extracted quantum capacitance ($C_q \propto DOS$) of the 6-layer BP sample perfectly fits the exponential relationship ($DOS = \alpha \cdot \exp(\frac{|E|-|E_b|}{E_U})$ with $E_F$ near the band edge, and $E_F$ can be obtained through the charge conservation relationship $E_F = e \cdot \int_0^{V_g} (1-\frac{C_t}{C_g})dV_g$ [15, 30-32]. The fittings (denoted by the red line in Fig. 3b) directly yield the bandgap size $E_g = E_{bc} - E_{bv} = 0.68$ eV and the band tail widths $E_U = 60$ meV and 50 meV for the valence and conduction bands respectively. Here $\alpha$ is a fitting constant. To exclude the influence of charge trap states, we performed capacitance measurement using a high excitation frequency (~100 kHz). The sample temperature was in the range of 20 K-30 K (kT<<$E_U$) in order to generate enough thermal electrons injected into the conduction band. This guaranteed the measurement accuracy in determining $E_{bc}$ and the band tail width of the conduction band.

Based on the analysis discussed above, we obtained the band gap widths of monolayer (Fig. 3d) and bilayer phosphorene (Fig. 3e). Detailed analysis can be found in the Supporting Information. BP exhibits a tunable bandgap from ~1.4 eV (monolayer) to ~0.3 eV (bulk) which decreases monotonically with the sample thickness (Fig. 3f). These results are consistent with theoretical predictions[2, 18, 19, 34] and experimental results measured from phosphorene field effect transistors[16]. We noted that the measured bandgap size in our monolayer BP is smaller than previous optical results ~2 eV[8]. This could be due to the oxidation of BP as demonstrated in the following theoretical calculations. We further identified the band tail width ($E_U \sim 100$ meV) in monolayer phosphorene which is in the same order compared to that obtained in monolayer MoS$_2$[27], indicating a high density of impurities and disorders in BP samples.

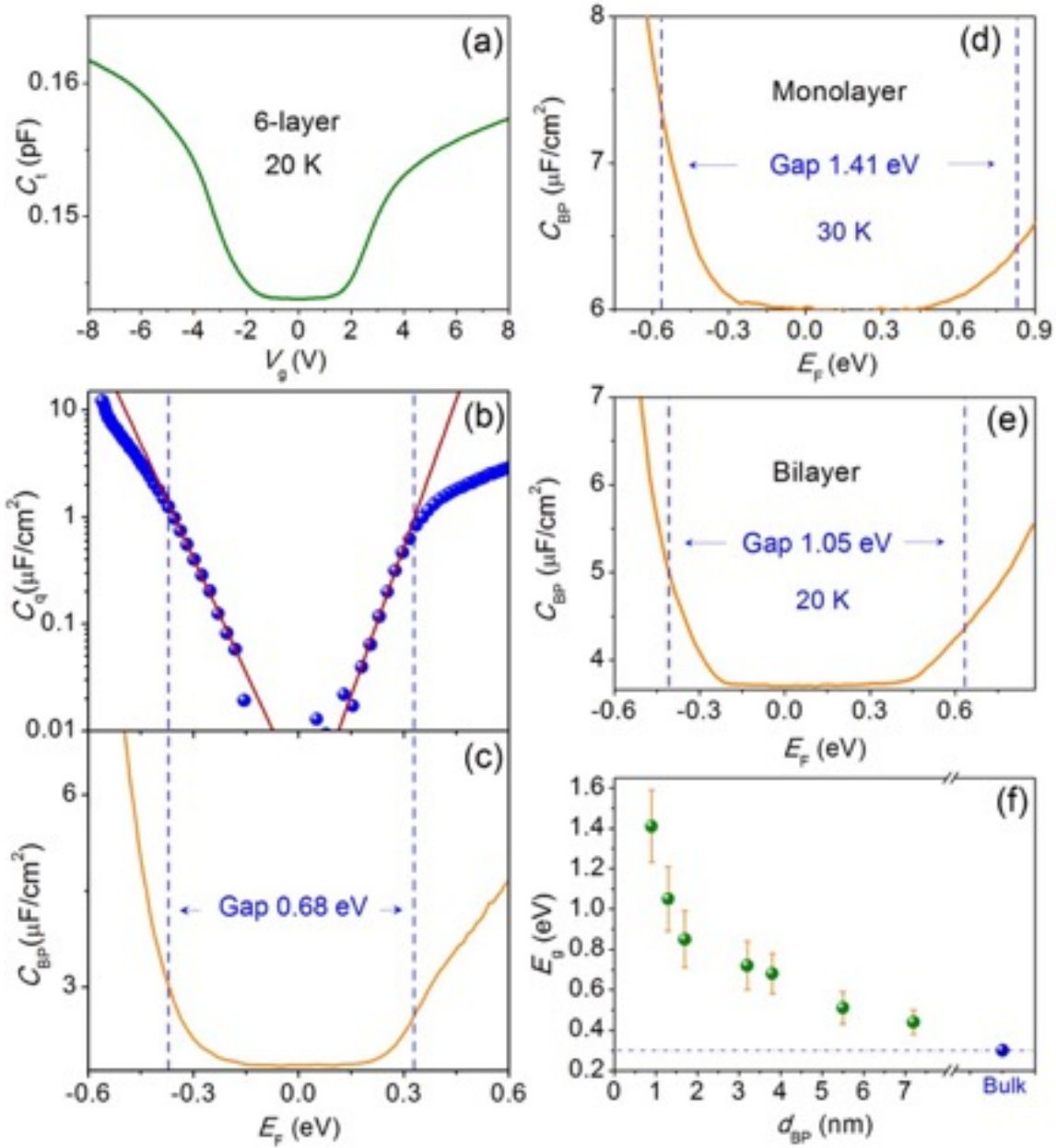

**Figure 3.** Characterization of the band tail and bandgap width of thin BP layers. (a-c) Total capacitance $C_t$ (a), quantum capacitance $C_q$ (b) and $C_{BP}$ (c) as a function of Fermi energy measured in the 6-layer BP sample at 20 K with an excitation voltage of 200 mV and a high excitation frequency of 100 kHz. The red solid line in (b) is the fitting based on

$$DOS = \alpha \cdot \exp(\frac{|E|-|E_b|}{E_U})$$

. (d, e) $C_{BP}$ measured as a function of Fermi energy in monolayer (d) and bilayer (e) phosphorene with an excitation voltage of 400 mV. The blue dashed lines denote the positions of conduction and valence bands of BP determined by fitting the experimental data. (f) The bandgap width of thin BP layers plotted as a function of thickness. The blue dot denotes the bandgap width of bulk BP.

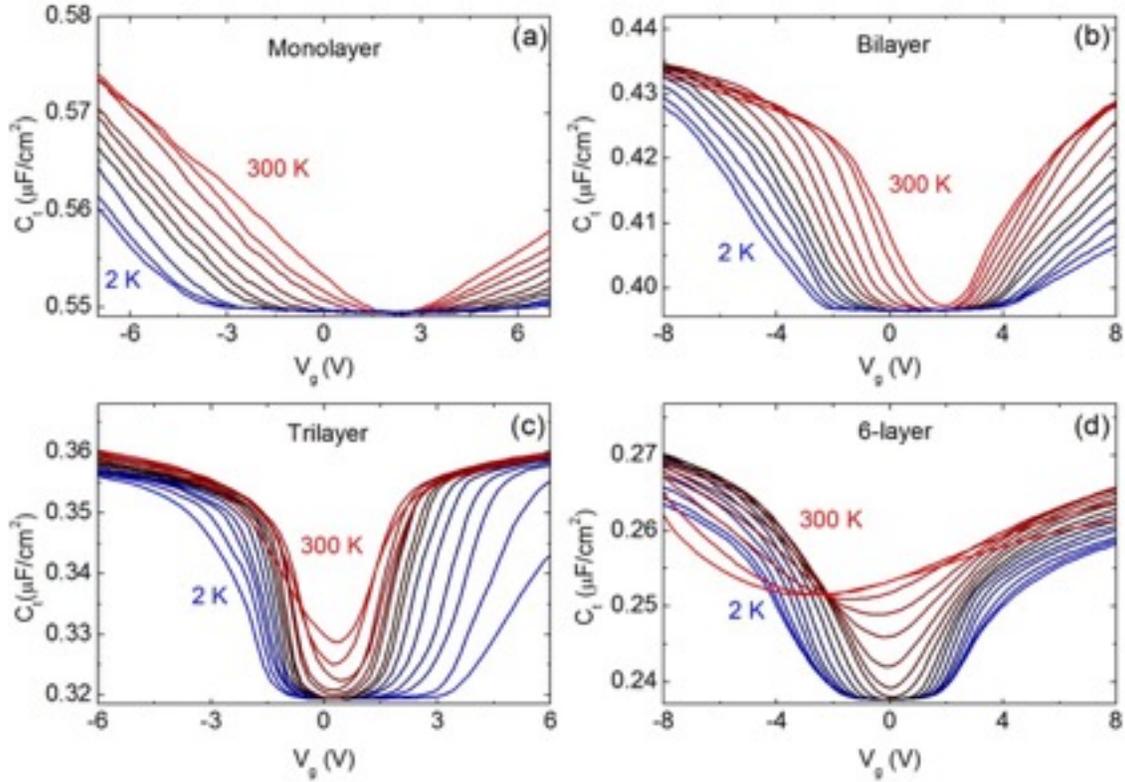

**Figure 4.** Temperature-dependent capacitance in monolayer and few layer phosphorene. (a-d) Capacitance $C_t$ measured as a function of gate voltage $V_g$ in monolayer (a), bilayer (b), trilayer (c) and 6-layer (d) phosphorene at different temperatures. The excitation voltage and frequency are 400 mV and 100 Hz for (a), 400 mV and 100 Hz for (b), 200 mV and 100 kHz for (c), and 100 mV and 100 kHz for (d) respectively.

One interesting and unexpected feature we discovered in monolayer and few layer phosphorene is their significant change of electron state population detected by capacitance measurement at different temperatures. The measured capacitance unusually increases with increasing temperature (Fig. 4a-d) when the Fermi energy is inside the band gap. Such a phenomenon becomes more pronounced for thick BP samples. For example, in the 6-layer sample, the gap looks almost "closed up" (Fig. 5b, e) around 150 K ($kT \sim 20$ meV). Further increasing of temperature leads to a large increase of $DOS \sim 1\times10^{13} \text{eV}^{-1}\text{cm}^{-2}$ inside the bandgap (Fig. 5c, f). Since thermal excitation of intrinsic electron states of BP is not able to cause this giant change of electron state population, we propose the following mechanism based on charge trapping effects to address this interesting phenomenon in few-layer BP. The charge trapping effects[3, 35] can be detected by performing positive and negative gate sweeping experiment. When the sample temperature is higher than 200K, a pronounced hysteresis effect is observed (see Fig. 5a-c and Fig. S4 in the Supporting Information). Similar to previous transport measurements[3, 35], a positive hysteresis direction is the sign of charge trapping effect which is different from capacitive coupling to the BP.[36] At low temperatures, the hysteresis effects disappear, further supporting the charge trapping mechanism since charge traps are frozen. We believe that the charge trap states are distributed inside the bandgap till the band edges of BP (Fig. 5g). At low temperatures, charge traps are suppressed and hence a finite gap structure is observed, while at high temperatures, charge trap states significantly contribute to the measured capacitance and increase the DOS. At a sufficiently high temperature, the gap looks "closed up" due to remarkable contribution of thermally excited charge trap states.

To quantitatively verify the density of charge traps $D_{it}$, we performed capacitance measurement at low temperatures (~2 K) with different excitation frequencies $f$. This is because the charge traps should have a relaxation time $\tau_{it}$. At a high excitation frequency $1/f < \tau_{it}$, the relaxation of charge trap states is suppressed, while when $1/f > \tau_{it}$, the charge trap states are excited and contribute to the measured quantum capacitance by adding an additional parallel capacitance $e \cdot D_{it}$. Hence, the charge trap densities can be expressed by $D_{it} = (C_{q\_lowf} - C_{q\_highf})/e$ [15]. The extracted charge trap densities in monolayer and few layer BP samples are in the range of $10^{12} - 10^{13}$ eV$^{-1}$cm$^{-2}$ (see Fig. 5h-j). Because of the limitation of the frequency range (100 Hz~1 MHz) in our experiment, $D_{it}$ could be underestimated. The charge trap densities in monolayer and bilayer phosphorene are only shown for the hole carrier side, as the presence of the Schottky barrier at electron side could lead to an inaccuracy in determining $D_{it}$ at 2 K. The observed large trap densities should not come from the insulating layer of hBN since hBN has been proven to be ultra-smooth and disorder-free dielectric materials without introducing charged impurities. We believe that the charge traps in BP are mainly due to the impurities and disorders induced in the fabrication processes particularly from the surface degradation caused by oxygen and water moisture. Previous transport results have shown that the on-off ratio in BP field-effect transistors (FETs) is small at high temperatures[1-3, 37]. Obviously, the excited charge traps should contribute a lot when measuring the on-off ratio based on conventional FET device configuration. In capacitance measurements, the DOS displays obvious asymmetry in electron and hole sides near the band edges of BP. A similar feature was also reported in previous transport measurements[3] which was attributed to the difference of electron and hole injection rates. Our capacitance results indicate that larger on-off ratio could be achieved in monolayer and bilayer phosphorene field-effect transistors[16] at room temperature.

This is because the bandgap has not been fully "closed up" in monolayer and bilayer phosphorene at room temperature (Fig. 4a, b).

In addition to charge trap effects, strain induced deformation could be a possible reason for the bandgap variation in few-layer BP. It is known that strain induced deformation in the direction perpendicular to the phosphorene plane could lead to a semiconductor-to-metal transition[18, 34]. Theoretical calculations[18, 34] predict that the compression in phosphorene can reduce the bandgap width and eventually lead to band crossings if the deformation is sufficiently large. In our vertical heterostructures, the only factor that could induce strain in BP is the thermal expansion of BP at high temperatures. However, the thermal expansion coefficient of BP[38] is $\sim 13 \times 10^{-6}$ K$^{-1}$ which causes only a very limited strain effect.

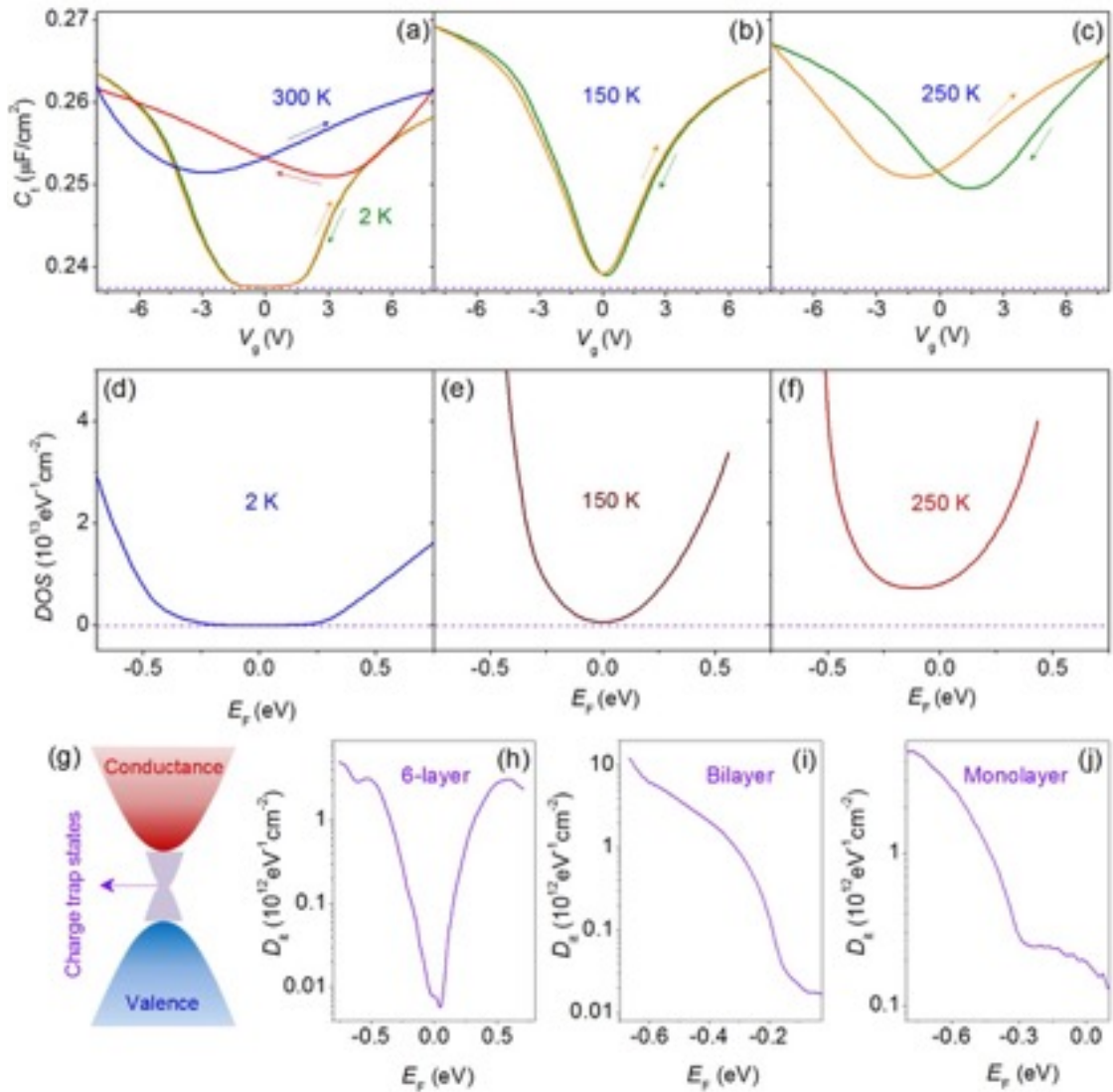

**Figure 5.** Temperature-dependent population of electron states in BP. (a-c) Hysteresis of measured capacitance $C_t$ of the 6-layer BP sample at different temperatures. (d-f) DOS of the 6-layer BP sample at different temperatures. (g) Schematic diagram showing the distribution of charge trap states inside the bandgap and near the band edges. (h-i) Charge trap densities in 6-layer (h), bilayer (i) and monolayer (j) BP samples.

To understand the quality degradation of BP and its influence on the bandgap size under atmospheric conditions, we performed the first-principles calculation on BP bonded with oxygen (O) and hydroxide ions (OH) based on density functional theory (DFT) method using the VASP code, within the projector augmented wave (PAW) method[39]. The exchange and correlation effects are described by Perdew-Burke-Ernzerhof (PBE) form of the generalized gradient approximation. The Monkhorst-Pack k point-mesh is chosen to be 5×1×5 for monolayer BP supercell models and the cutoff energy is set to 550 eV. The total ground state energy is converged within $10^{-6}$ eV per unit cell, and all the atoms are fully relaxed until the Hellmann-Feynman forces exerted on each atom become less than 0.01 eV/Å. The spin polarization has been taken into account in all of our calculations. As shown in Table 1, the calculated bandgap size for pristine monolayer phosphorene is 0.904 eV, which is consistent with previous PBE results[24]. Although the PBE method might underestimate the bandgap size of BP, it provides useful information on the relative change of bandgap of phosphorene with defects. Geometric configurations of O and OH defects of different concentrations in phosphorene are shown in Fig. 6. Table 1 illustrates the calculation results for such O and OH adatoms bonded on the surface of monolayer BP, including bonding lengths and bonding angles. O slightly increases the bandgap size of monolayer phosphorene, whereas OH can significantly decrease its bandgap size. Thus the oxidation effect could effectively lead to a smaller bandgap in monolayer phosphorene, through the modification of electronic properties by the extrinsic point defects.

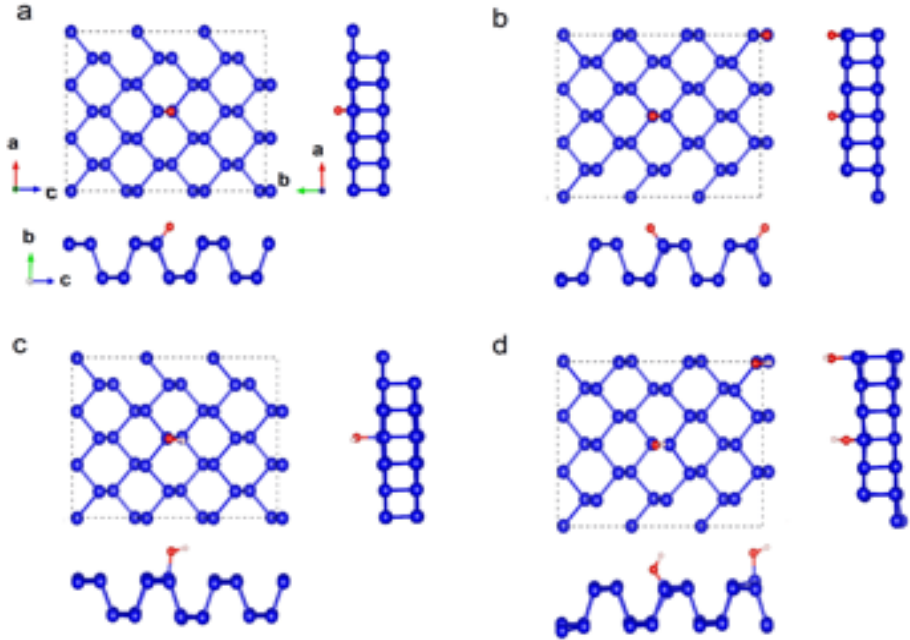

**Figure 6.** Geometric configurations of (a) one O, (b) two O atoms, (c) one OH and (d) two OH defects adsorbed on the surface of monolayer phosphorene.

**Table 1.** Calculated structural parameters and bandgap of monolayer phosphorene.

| Structure | P-O (Å) | O-H (Å) | P-O/plane (°) | P-O-H (°) | Band Gap (eV) |
|---|---|---|---|---|---|
| BP-pristine | -- | -- | -- | -- | 0.904 |
| BP-1O | 1.506 | -- | 52.535 | -- | 1.035 |
| BP-2O | 1.504 | -- | 48.802 | -- | 1.133 |
|  | 1.504 |  | 52.457 |  |  |
| BP-1OH | 1.684 | 0.978 | 88.042 | 108.709 | 0.131 |
| BP-2OH | 1.630 | 0.980 | 123.337 | 113.548 | 0.410 |
|  | 1.706 | 0.975 | 89.155 | 107.833 |  |

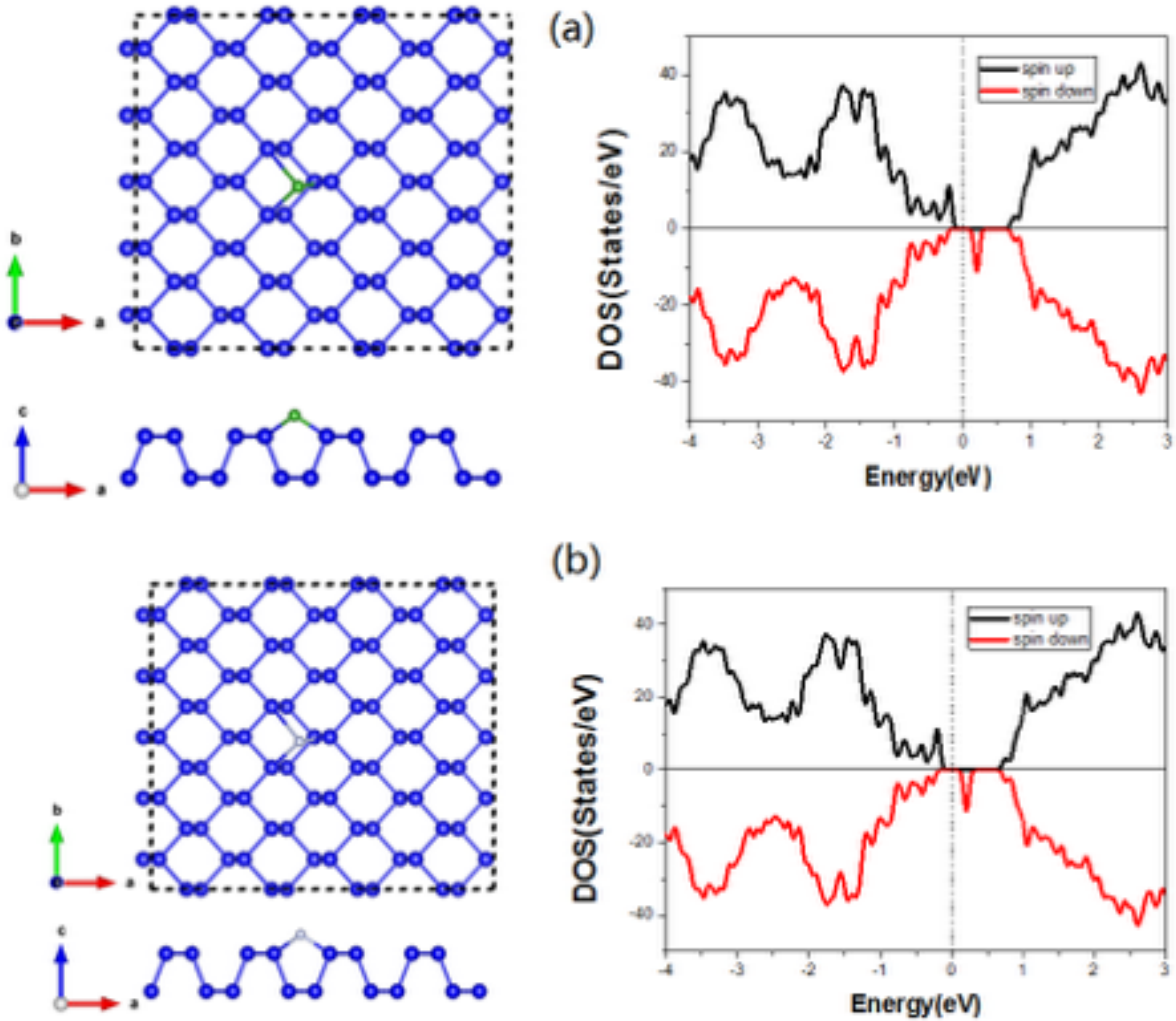

**Figure 7.** Geometric configurations and calculated DOS of (a) B atom, and (b) N atom, adsorbed on the surface of monolayer phosphorene.

To investigate the influence of the B and N defects which possibly come from hBN in the heterostructure. We employed a supercell containing 80 P atoms to simulate B and N defects in the BP, as illustrated in Fig. 7. In contrast to O absorption, for B and N adatoms (either adsorbed on BP surface or interstitially embedded in BP layer) spin splitting of DOS peaks near valence band edge can be found, consequently the spin polarized charge density is localized on the atoms around the adatoms. Thus, these impurities might provide additional sources of charge trap states discovered in our experiments. As shown in Fig. S5 of Supporting Information, four substitution doping configurations are also considered, including B, N doing and B/N codoing with two different B-N defect distances. We found all these defects in BP generally decrease the band gap of BP. For the N defect (having same valence as P) in BP, the band gap is reduced to 0.79 eV without distinct mid-gap state near conduction band. For the other defects, the band gaps decrease considerably with the appearance of middle states, as a result of hole doping owing to the B defect in the BP.

## 3. Conclusions

In summary, the vertical BP/hBN heterostructure configuration offers great advantages in probing the electron states of monolayer and few layer phosphorene at temperatures down to 2 K for capacitance spectroscopy. Important information such as DOS, bandgap width and charge trap density has been obtained from quantum capacitance measurement. The thermal excitation of charge trap states has been identified to be the main reason for the giant temperature-dependence of the electron state population, capacitance and transport hysteresis and asymmetry distribution of DOS. Combined with first principles calculations, our experimental results show directions for improving the quality and on-off ratios of BP field-effect transistors.

**AUTHOR INFORMATION**

**Corresponding Author**
*E-mail: phwang@ust.hk.

**Acknowledgements**

Financial support from the Research Grants Council of Hong Kong (Project Nos. 16302215, HKU9/CRF/13G, 604112 and N_HKUST613/12) and technical support of the Raith-HKUST Nanotechnology Laboratory for the electron-beam lithography facility at MCPF (Project No. SEG_HKUST08) are hereby acknowledged. This work was also supported by the National Key Basic Research Program of China (Grant No. 2015CB921600), the National Natural Science Foundation of China (NSFC, Grant No. 11274222), the QiMingXing Project (Project No. 14QA1402000) of the Shanghai Municipal Science and Technology Commission, the Eastern Scholar Program, and the Shuguang Program (Grant No. 12SG34) from the Shanghai Municipal Education Commission.